\documentclass[twocolumn,twoside,slac_two]{revtex4}
\usepackage{graphicx}
\usepackage{fancyhdr}
\begin{document}

\title{Radiation from accelerated particles in relativistic jets with shocks, shear-flow, and reconnection}

%

\author{K.-I. Nishikawa}
\affiliation{Center for Space Plasma and Aeronomic Research,
University of Alabama in Huntsville, 320 Sparkman Drive, ZP12,
Huntsville, AL 35805, USA}

\author{P. Hardee}
\affiliation{Department of Physics and Astronomy,
  The University of Alabama, Tuscaloosa, AL 35487, USA}

\author{B. Zhang}
\affiliation{Department of Physics, University of Nevada, Las
Vegas, NV 89154, USA}

\author{I. Du\c{t}an}
\affiliation{Institute of Space Science, Atomistilor 409, Bucharest-Magurele RO-077125, Romania}

\author{M. Medvedev}
\affiliation{Department of Physics and Astronomy, University of Kansas, KS
66045, USA}

\author{E. J. Choi}
\author{K. W. Min}
\affiliation{Korea Advanced Institute of Science and Technology, 
Daejeon 305-701, South Korea}

\author{J. Niemiec}
\affiliation{Institute of Nuclear Physics PAN, ul. Radzikowskiego 152, 31-342 Krak\'{o}w, Poland}

\author{Y. Mizuno}
\affiliation{Institute of Astronomy
National Tsing-Hua University,
Hsinchu, Taiwan 30013, R.O.C}

\author{A. Nordlund}
\author{J. T. Frederiksen}
\affiliation{Niels Bohr Institute, University of Copenhagen, 
Juliane Maries Vej 30, 2100 Copenhagen \O, Denmark}

\author{H. Sol}
\affiliation{LUTH, Observatore de Paris-Meudon, 5 place Jules Jansen, 92195 Meudon Cedex, France}

\author{M. Pohl}
\affiliation{Institue of Physics and Astronomy, University of Potsdam, Karl-Liebknecht-Strasse 24/25
14476 Potsdam-Golm, Germany} 

\author{D. H. Hartmann}
\affiliation{Department of Physics and Astronomy, Clemson University, Clemson, SC 29634, USA}

\begin{abstract}
We have investigated particle acceleration and shock structure associated with an unmagnetized relativistic jet propagating into an unmagnetized plasma. Strong magnetic fields generated in the trailing jet shock lead to transverse deflection and acceleration of the electrons. We have self-consistently calculated the radiation from the electrons accelerated in the turbulent magnetic fields. We find that the synthetic spectra depend on the bulk Lorentz factor of the jet, the jet temperature, and the strength of the magnetic fields generated in the shock. We have also begun study of electron acceleration in the strong magnetic fields generated by kinetic shear (Kelvin-Helmholtz) instabilities. Our calculated spectra should lead to a better understanding of the complex time evolution and/or spectral structure from gamma-ray bursts, relativistic jets, and supernova remnants.
\end{abstract}

\maketitle

\thispagestyle{fancy}


\section{INTRODUCTION}
Recent kinetic simulations have focused on magnetic field generation via electromagnetic plasma instabilities in unmagnetized flows without velocity shears.
Three-dimensional (3D) particle-in-cell (PIC) simulations of Weibel turbulence [Nishikawa et al. 2005, 2009a]
have demonstrated subequipartition magnetic field generation.
We have calculated, self-consistently, the radiation from electrons accelerated in
the turbulent magnetic fields. We found that the synthetic spectra depend on the
Lorentz factor of the jet, the jet's thermal temperature, and the strength of the generated
magnetic fields \citep{nishikawa11,nishikawa12}. 

Velocity shears also must be considered when studying particle acceleration scenarios, since these trigger the kinetic Kelvin-Helmholtz instability (KKHI). In particular the KKHI has been shown to lead to particle acceleration and magnetic field amplification in relativistic shear flows [Alves et al. 2012; 
Liang et al. 2012]. Furthermore, a shear flow upstream of a shock can lead to density inhomogeneities via
the MHD Kelvin-Helmholtz instability (KHI) which may provide important scattering
sites for particle acceleration.

\vspace*{-0.cm}
\begin{figure*}[ht]
\begin{minipage}[t]{100mm}
\resizebox{0.95\hsize}{!}{\includegraphics[width=12cm]{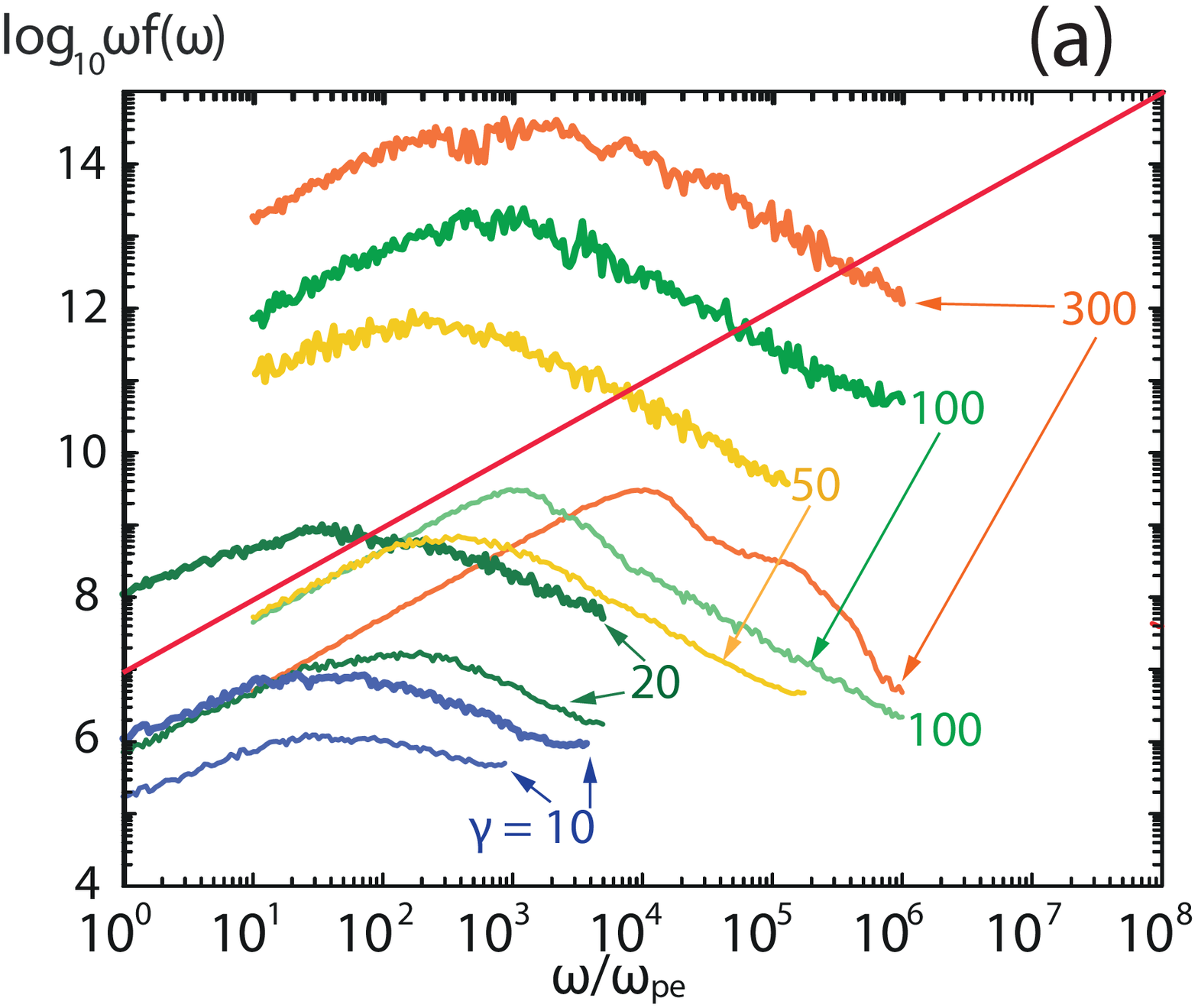}
\,
\qquad 
\includegraphics[width=10cm]{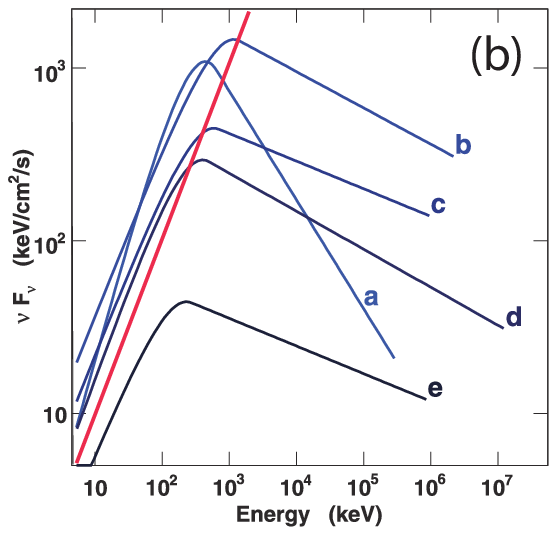}} 

\end{minipage}
\begin{minipage}[t]{60mm}
\vspace*{-4.2cm}
\caption{\baselineskip 12pt  Figure \ref{spect}a shows the spectra for the cases of $\gamma =$ 10, 20, 50, 100, and 300  with cold (thin lines) and warm (thick lines) electron jets.    Figure \ref{spect}b shows modeled Fermi spectra in $\nu F_{\nu}$ units at early (a) to late (e) times [Abdo et al. 2009]. The red lines indicate slope in $\nu F_{\nu} \sim 1$ \label{spect}} 
\end{minipage}
\end{figure*}

\section{THE STANDARD RADIATION MODEL}
A synchrotron shock model has been widely adopted as describing the
radiation mechanism thought responsible
for observed broad-band GRB afterglows [e.g., Zhang
\& Meszaros 2004; Piran 2005a; Zhang 2007, Nakar 2007].  Due to the lack of a first principles theory of
collisionless shocks, a purely phenomenological approach to the
model of afterglow radiation has been prescribed. Firstly, electrons are assumed
to be ``Fermi'' accelerated at relativistic shocks and to have a
power-law distribution with index $p$, where $N(E_{\rm e})dE_{\rm e} \propto E^{-p} dE_{\rm e}$. 
Secondly, a fraction $\xi_{\rm e}$ (generally taken to be
$\lesssim 1$) of the electrons  are assumed to be
accelerated, and the total electron energy is assumed to be a fraction $\epsilon_{\rm e}$ 
of the total internal energy in the shocked region. Thirdly, the
strength of the magnetic field in the shocked region is unknown, but
its energy density ($B^{2}/8\pi$) is assumed to be a fraction
$\epsilon_{B}$ of the internal energy.  These assumed ``micro-physics''
parameters, $p, \xi_{\rm e}, \epsilon_{\rm e}$ and $\epsilon_{\rm B}$, whose values are
inferred from the spectral fits to the observations [e.g., Panaitescu \& Kumar 2001; Yost et al.\ 2003], 
reflect a lack of knowledge of the underlying microphysics [Waxman 2006].  It is our intent to place these parameters on a  firm physical basis.


\section{SELF-CONSISTENT RADIATION CALCULATION FROM PIC SIMULATIONS}
 Electrons are accelerated in the electromagnetic fields
generated by the Weibel and kinetic Kelvin-Helmholtz instabilities. Radiation can be calculated using the particle trajectories in
the self-consistent turbulent magnetic fields. This calculation includes Jitter radiation [Medvedev
2000, 2006] which is different from standard synchrotron emission. Radiation from electrons in our simulations is reported
in Nishikawa et al. [2011].

We have calculated the radiation spectra directly from our simulations by integrating the expression for the retarded power, derived from Li\'{e}nard-Wiechert potentials for a large number of representative particles in the
PIC representation of the plasma [Jackson 1999; Hededal 2005; Nishikawa et al. 2008a,b, 2009b, 2010,
2011; Martins et al. 2009; Sironi \& 
Spitkovsky 2009; Frederiksen et al. 2010]. Initially we verified the technique by calculating radiation from 
electrons propagating in a uniform parallel magnetic field [Nihikawa et al. 2009b].
It should be noted that spectra obtained from colliding jet simulations (fixed contact discontinuity) do not provide
spectra in the observer's rest frame, and cannot be compared with observed spectra [Sironi \& 
Spitkovsky 2009b].

The spectra shown in Figure \ref{spect}a are for emission from jets with Lorentz factors $\gamma =$ 10, 20, 50, 100, and 300  [Nishikawa et al. 2011, 2012]. In the figure we show two spectra for each Lorentz factor (represented by the same color line) for initially cold ($v_{\rm jet,th} = 0.01c$) (thin, lower lines) and initially warm ($v_{\rm jet,th} = 0.1c$) (thick, upper lines) jet electrons. 
Here the spectra are calculated for emission along the jet axis ($\theta = 0^{\circ}$).  The radiation shows 
a Bremsstrahlung-like spectrum  at low frequencies for the eleven cases (Hededal 2005) because the magnetic fields 
generated by the Weibel instability are rather weak and jet electron acceleration is modest. 
A low frequency slope of $\nu F_{\nu} = 1$ is indicated  by the straight red lines. 
The low frequency slopes in our synthetic spectra are very similar to those of the spectra in Fig. \ref{spect}b in Abdo et al.\ [2009], and show change with the Lorentz factor like the temporal evolution observed
by Fermi as shown in Fig. \ref{spect}b. Here we have not included radiation losses [Jaroschek et al. 2009; Medvedev  \& Spitkovsky 2009].





%
\vspace*{-0.0cm}
\begin{figure*}[ht!]
\resizebox{1.00\hsize}{!}{\includegraphics[width=18cm]{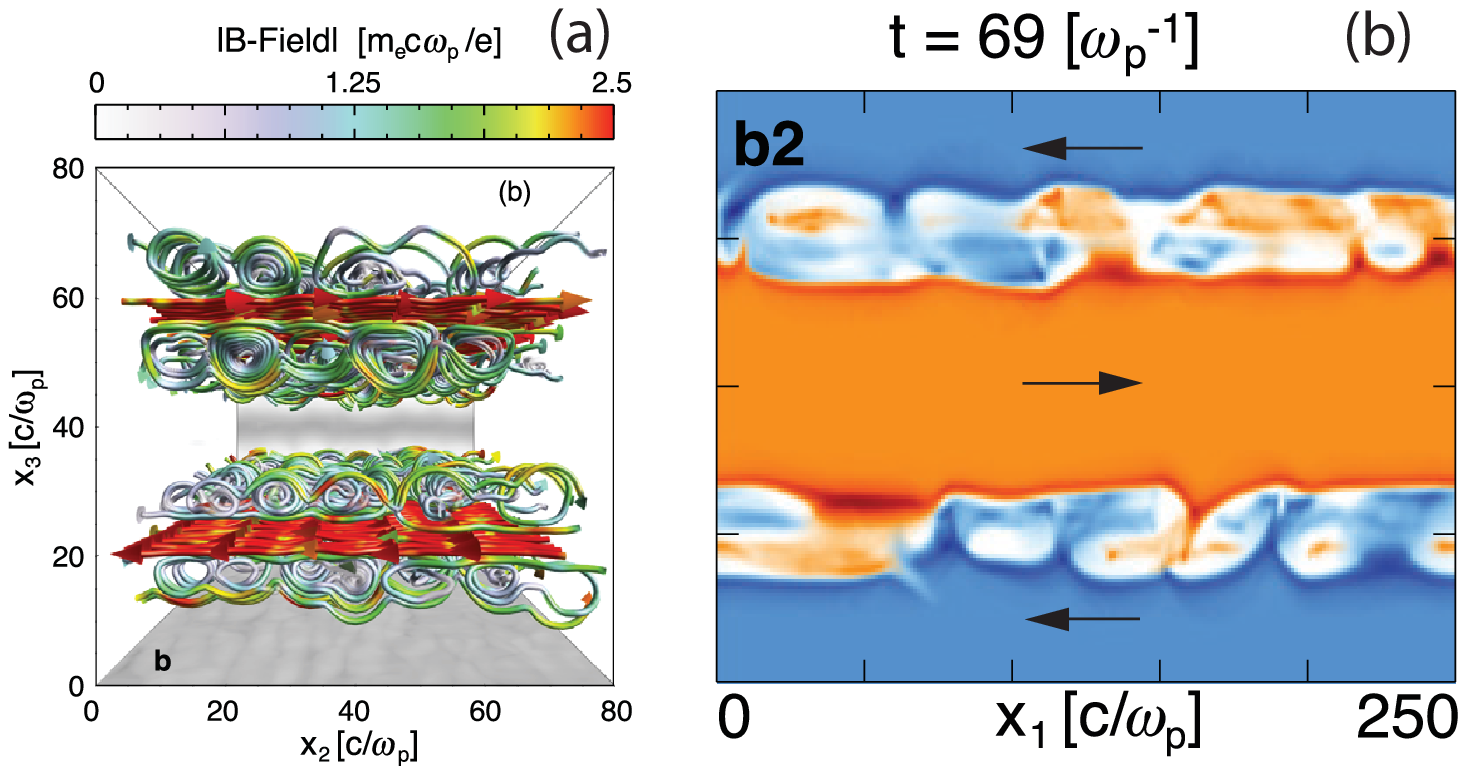}
\,
\qquad 
\includegraphics[width=12cm]{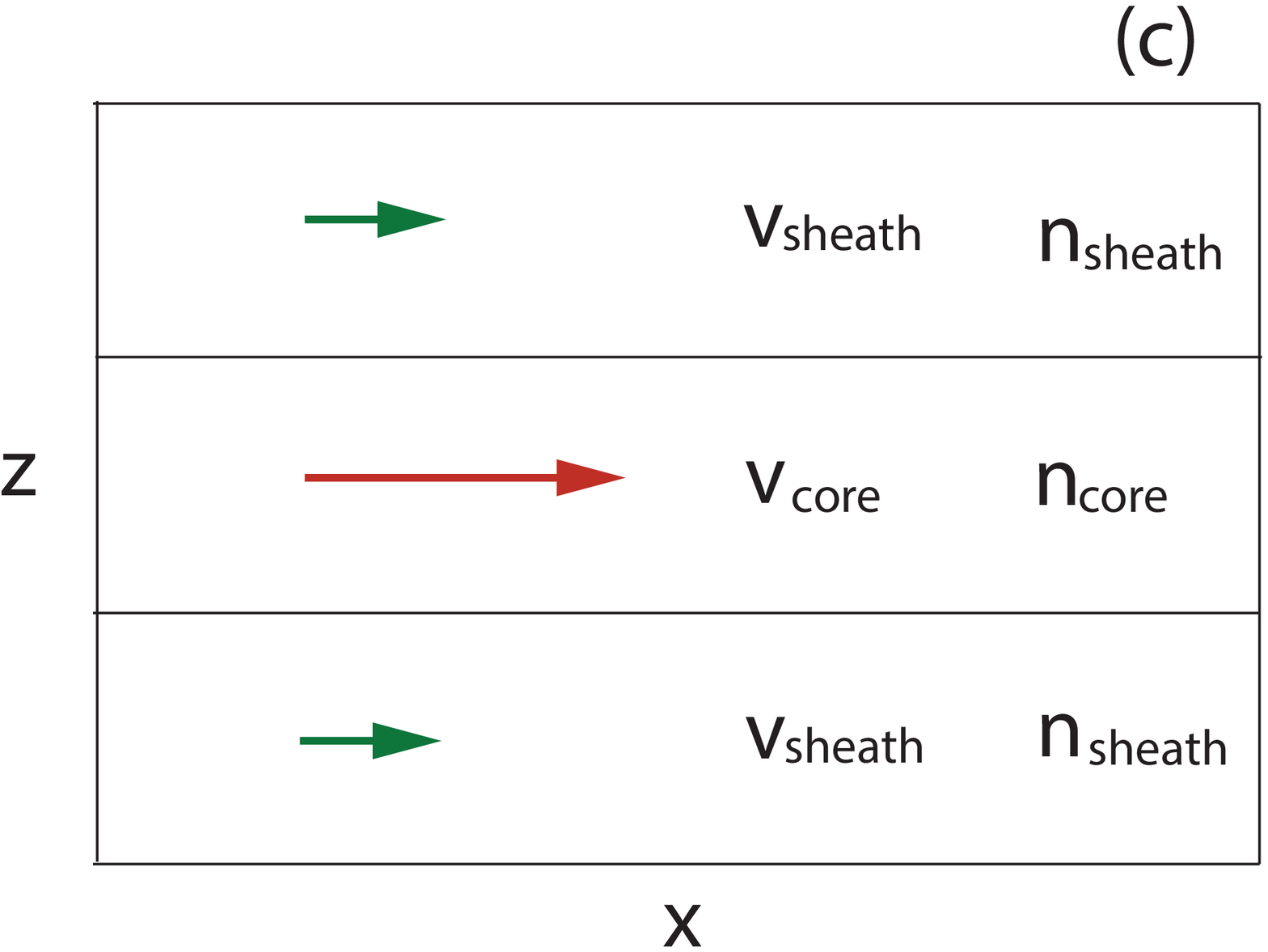}} 
\vspace*{-0.3cm}
\caption{\baselineskip 12.0pt Figure \ref{khiI}a shows magnetic field lines generated in the relativistic shear from Alves et al. [2012].  Figure \ref{khiI}b shows the electron density in orange (blue) of the plasma that flows in the positive (negative) $x_{\rm 1}$ direction. In Fig. \ref{khiI}b darker regions in the color map indicate high electron density, whereas lighter regions indicate low electron density.   Figure \ref{khiI}c  shows our simulation model where the sheath plasma can be stationary or moving in the same direction as the jet core. 
\label{khiI}}
\end{figure*}

\begin{figure*}[ht]
\resizebox{.75\hsize}{!}{\includegraphics[width=12cm]{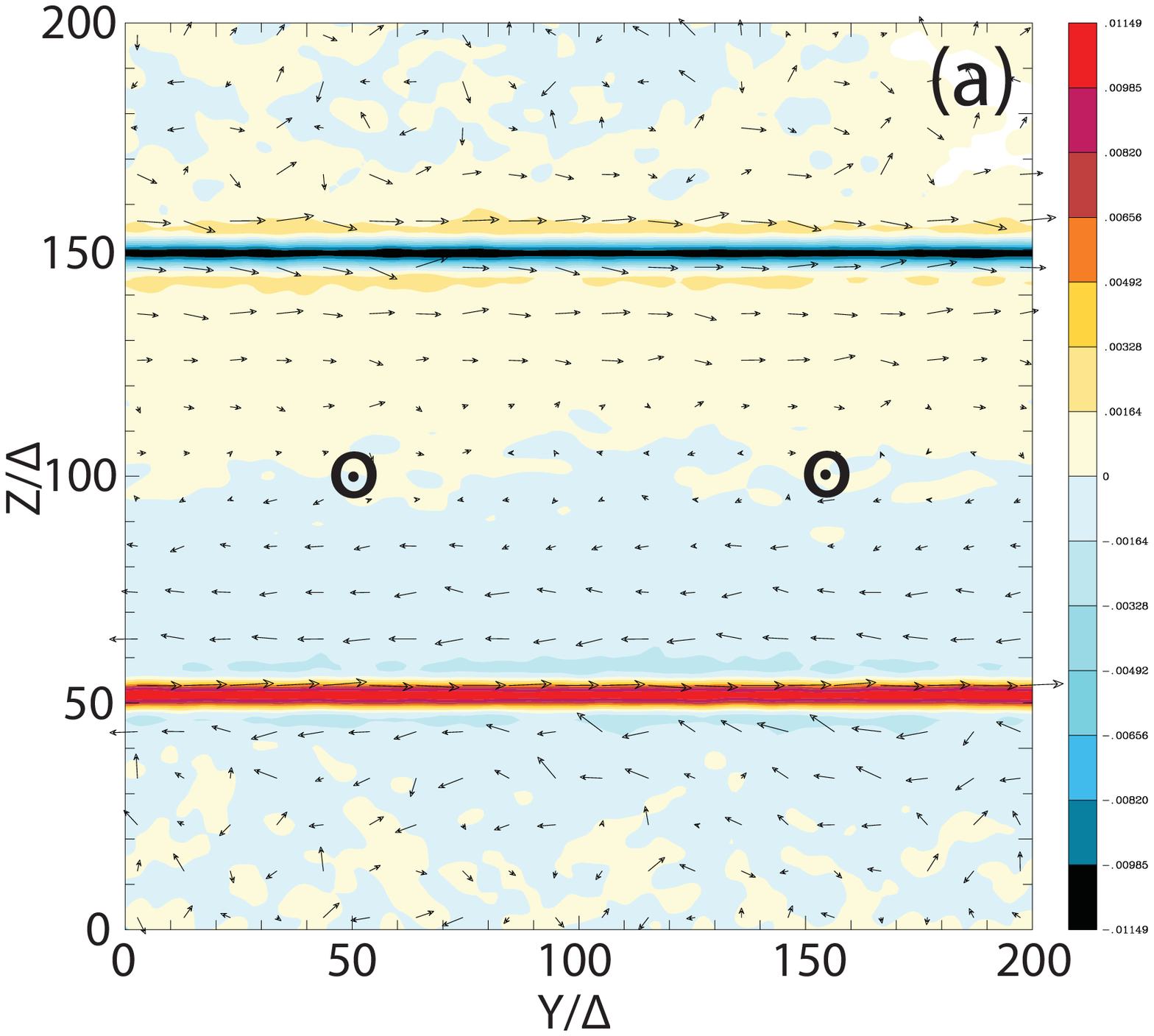}
\quad \quad \quad \quad \quad \quad \quad \quad \quad \quad \quad \quad \quad \quad \quad
\includegraphics[width=6.95cm]{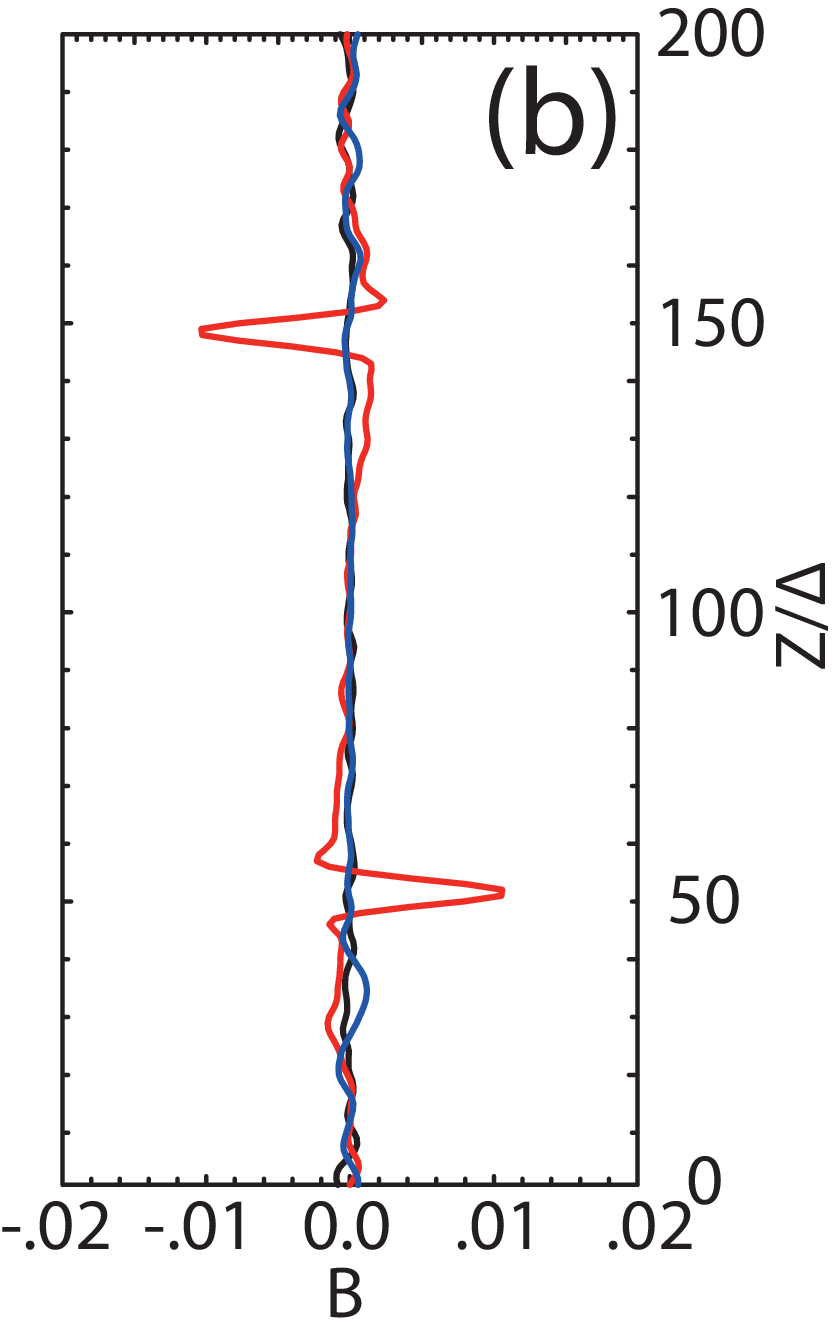} }
\,
\resizebox{.7560\hsize}{!}{\includegraphics[width=12cm]{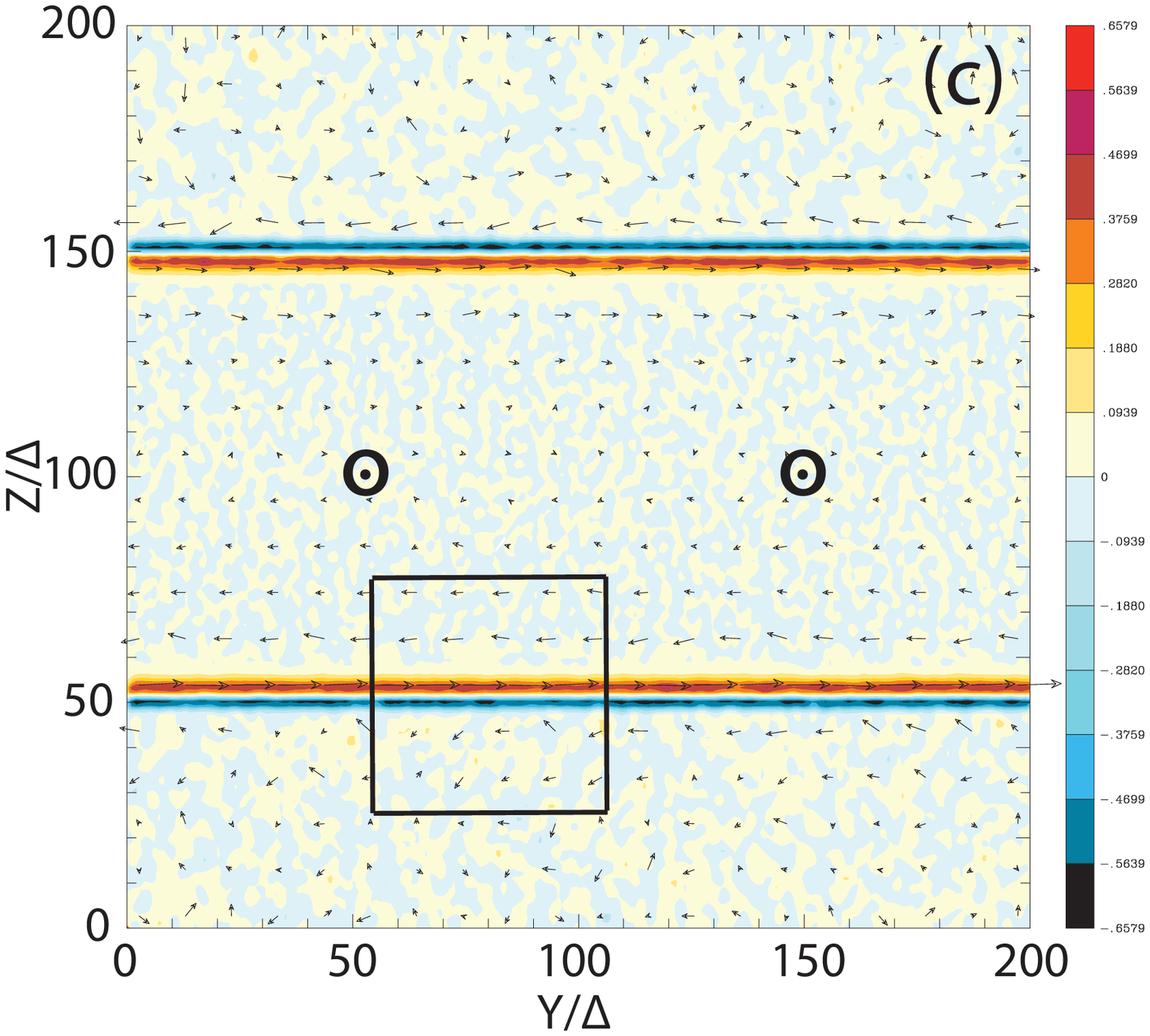}
\, 
\includegraphics[width=12cm]{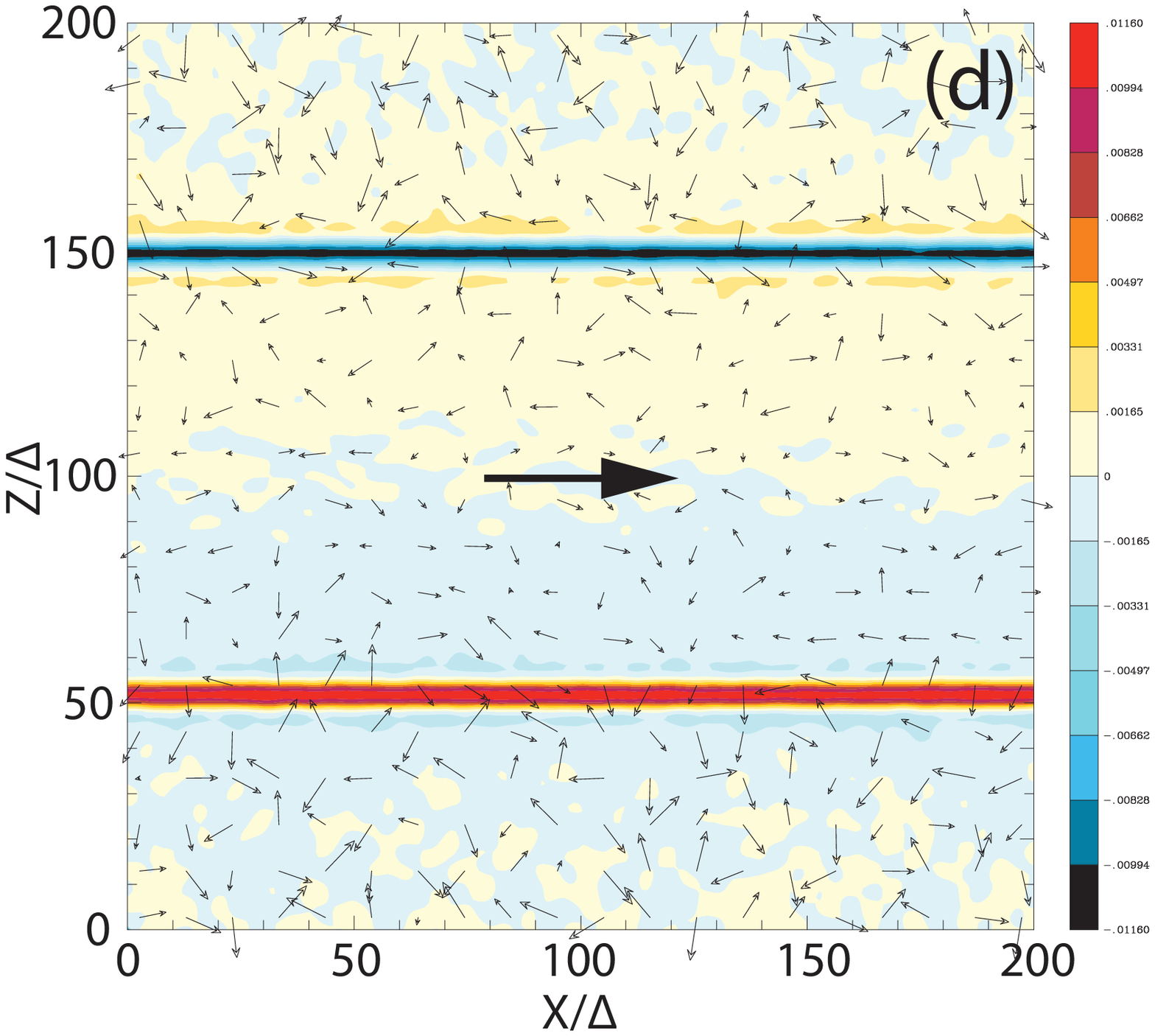}} 
\caption{Magnetic field structure generated by a relativistic electron-ion jet core
 with $\gamma = 15$ and stationary sheath plasma at simulation time $t = 70\,\omega_{\rm pe}^{-1}$.  The magnitude of $B_{\rm y}$  is plotted (a) in the  $y - z$ plane (jet flow out of the page) at the center of the simulation box, $x = 500\Delta$ (d) in the $x - z$ plane (jet flow in the $+x$-direction indicated by the arrow) at the center of the simulation box, $y = 100\Delta$.  Figure~\ref{kkhi}b shows $B_{\rm y}$ (red), $B_{\rm x}$ (black), and $B_{\rm z}$ (blue) at $x = 500\Delta$ and $y = 100\Delta$.
Figure~\ref{kkhi}c shows the $x$ component of the electric current (jet flow is out of the page). The current is positive on the core side and negative on the sheath side of the velocity shear. The positive current is stronger than the negative current, leading to $B_{\rm y}$ as shown in Figure~\ref{kkhi}b. The arrows show the magnetic field in the plane. \label{kkhi}}
\end{figure*}

\section{MAGNETIC FIELD GENERATION AND PARTICLE ACCELERATION BY THE KKHI}
\subsection{Initial Conditions and Previous Results}

Recently the kinetic KH instability (KKHI) has been investigated using  a relativistic counter-streaming velocity shear setup with $\gamma_{\rm 0} = 3$ [Alves et al.\  2012].  In this RPIC simulation the velocity shear occurs at the edges of  a velocity field with $v_{\rm 0}$  pointing in the positive  $x$ ($x_{\rm 1}$) direction  in the middle of the simulation box, with upper and lower quarters of the simulation box containing a velocity field with $v_{\rm 0}$ pointing in the negative $x_{\rm 1}$ direction as indicated by the arrows in Fig. \ref{khiI}c. Initially, the system was charge and current neutral.
The simulation box dimensions were
$250 \times 80 \times 80 (c/\omega_{\rm p})^3$,  where $\omega_{\rm p} = (4\pi n_{\rm e}^{2}/m_{\rm e})^{1/2}$ is the plasma frequency, with a resolution of 4 cells per $c/\omega_{\rm p}$. Periodic boundary conditions were applied in all directions.

Figure \ref{khiI}a shows  the magnetic field line topology for this relativistic case. Alves et al. found that  KKHI modulations are less noticeable in the relativistic regime because they are masked by a  strong DC magnetic field component  (negligible in the subrelativistic regime) with a magnitude greater than the AC component,.  As the amplitude of the KKHI modulations grows the electrons from one flow cross the shear-surfaces and enter the
counter-streaming flow. In their simulations the protons being heavier ($m_{\rm p}/m_{\rm e} = 1836$)  are unperturbed. DC current sheets which point in the direction of the proton velocity form around the shear-surfaces. These DC current sheets induce a DC component in the magnetic field shown in Fig. \ref{khiI}b.  The DC magnetic field is dominant in the relativistic scenario because a higher
DC current is set up by the crossing of electrons with a larger initial flow velocity and also because the growth rate of the AC dynamics is lower by $\gamma_{\rm 0}^{3/2}$ compared with a subrelativistic case.
It is very important to note  that this DC magnetic field is not captured in MHD [e.g., Zhang
et al. 2009) or fluid theories because it results from intrinsically
kinetic phenomena. Furthermore, since the DC field is stronger
than the AC field, a kinetic treatment is clearly required in order
to fully capture the field structure generated in unmagnetized or weakly magnetized
relativistic flows with velocity shear. This characteristic field
structure will also lead to a distinct radiation signature [Sironi \& Spitkovsky 2009b; Martins et al.
2009; Frederiksen et al. 2010; Nishikawa et al. 2011, 2012].

\begin{figure*}[ht]
\resizebox{.80\hsize}{!}{\includegraphics[width=18cm]{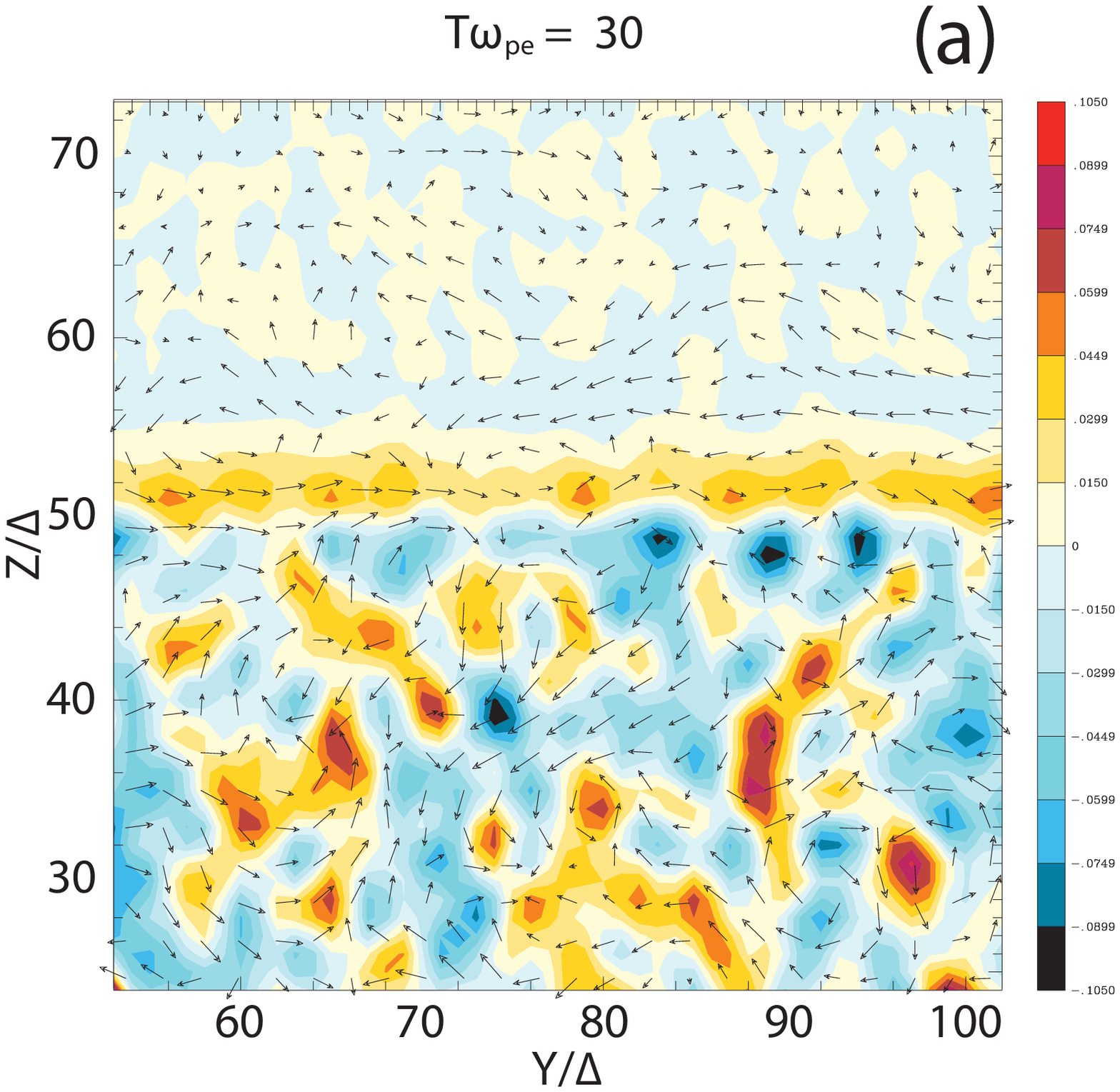}
\,
\includegraphics[width=18cm]{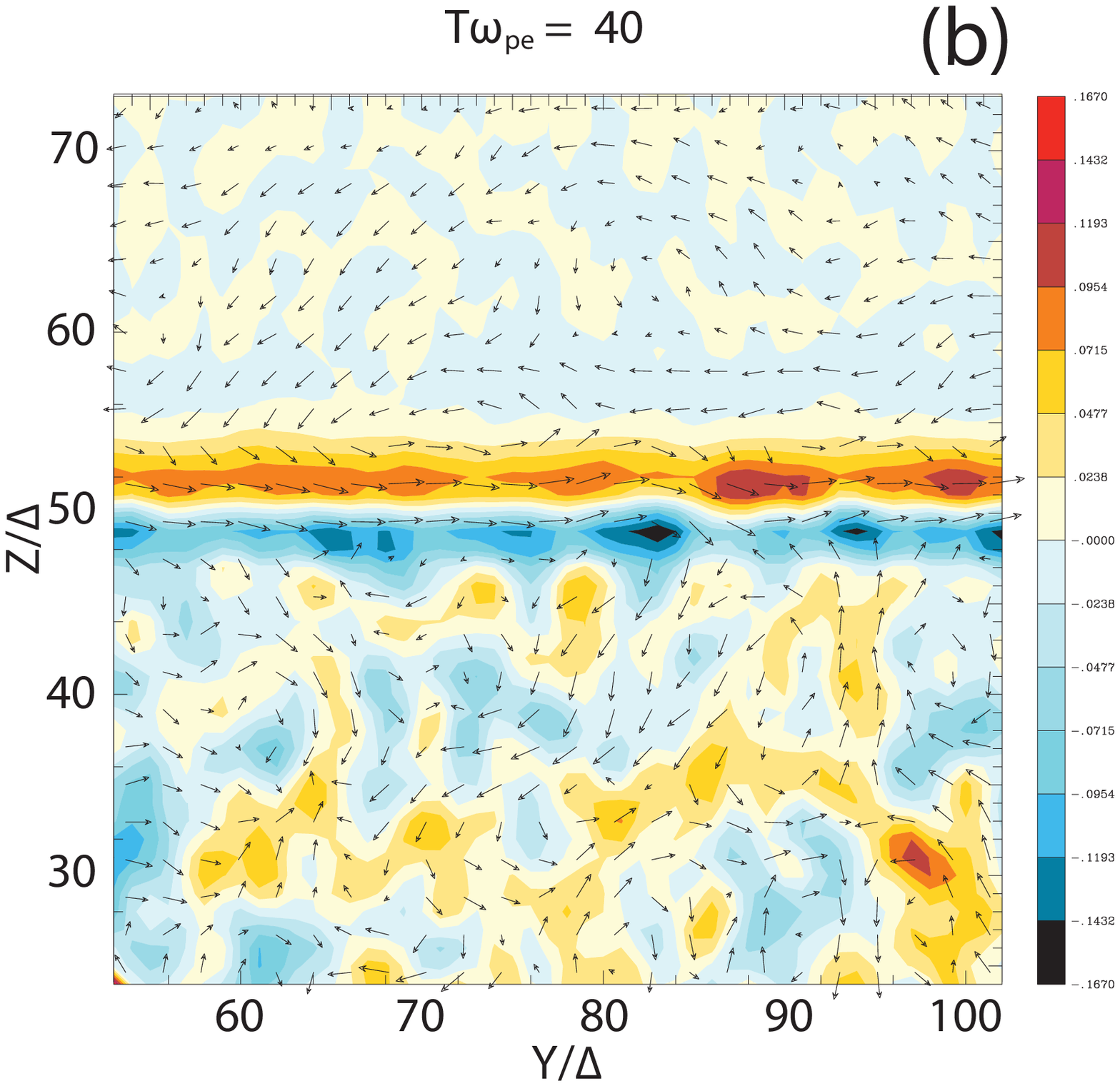} }
\,
\resizebox{0.80\hsize}{!}{\includegraphics[width=18cm]{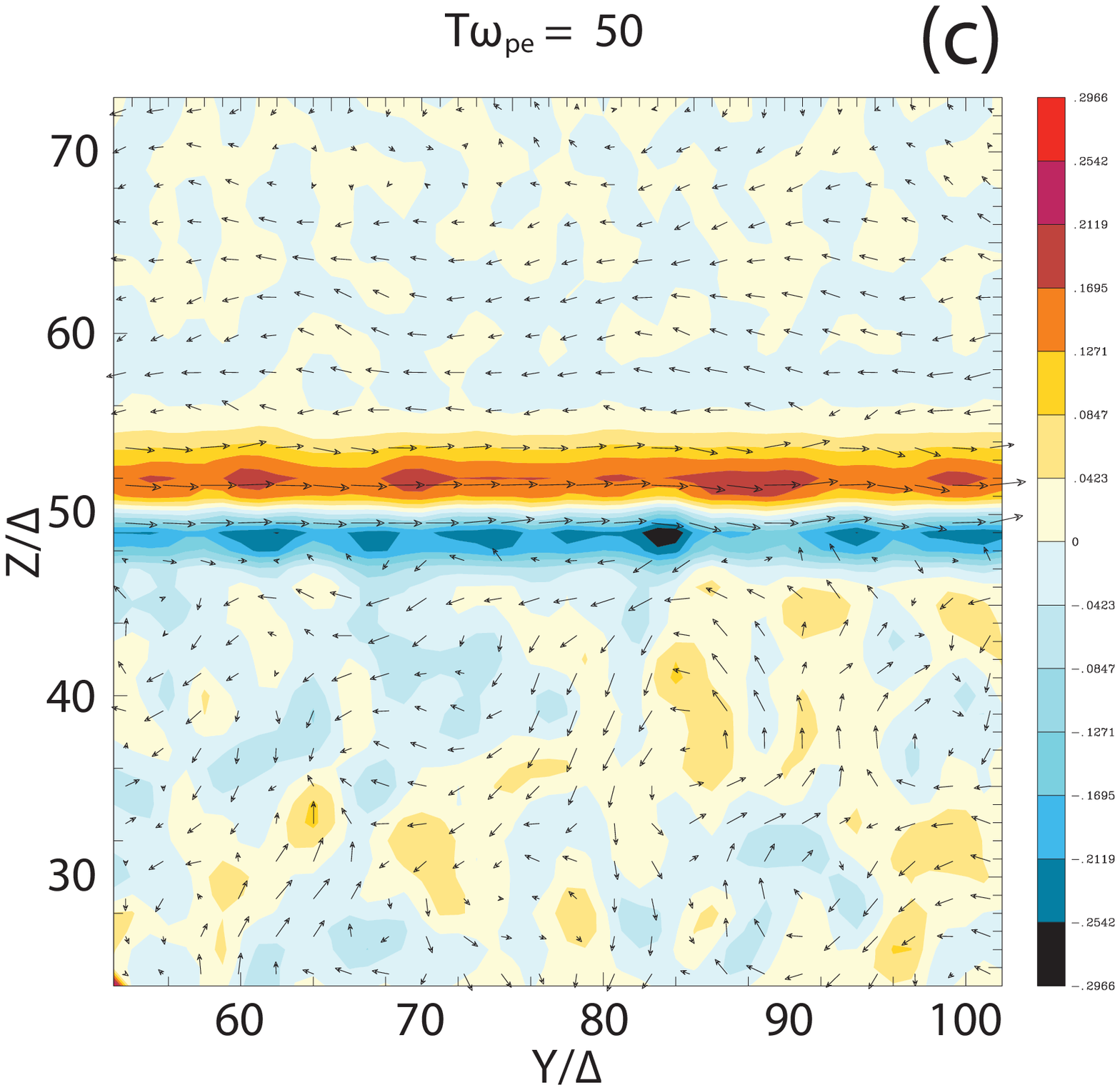}
\,
\includegraphics[width=18cm]{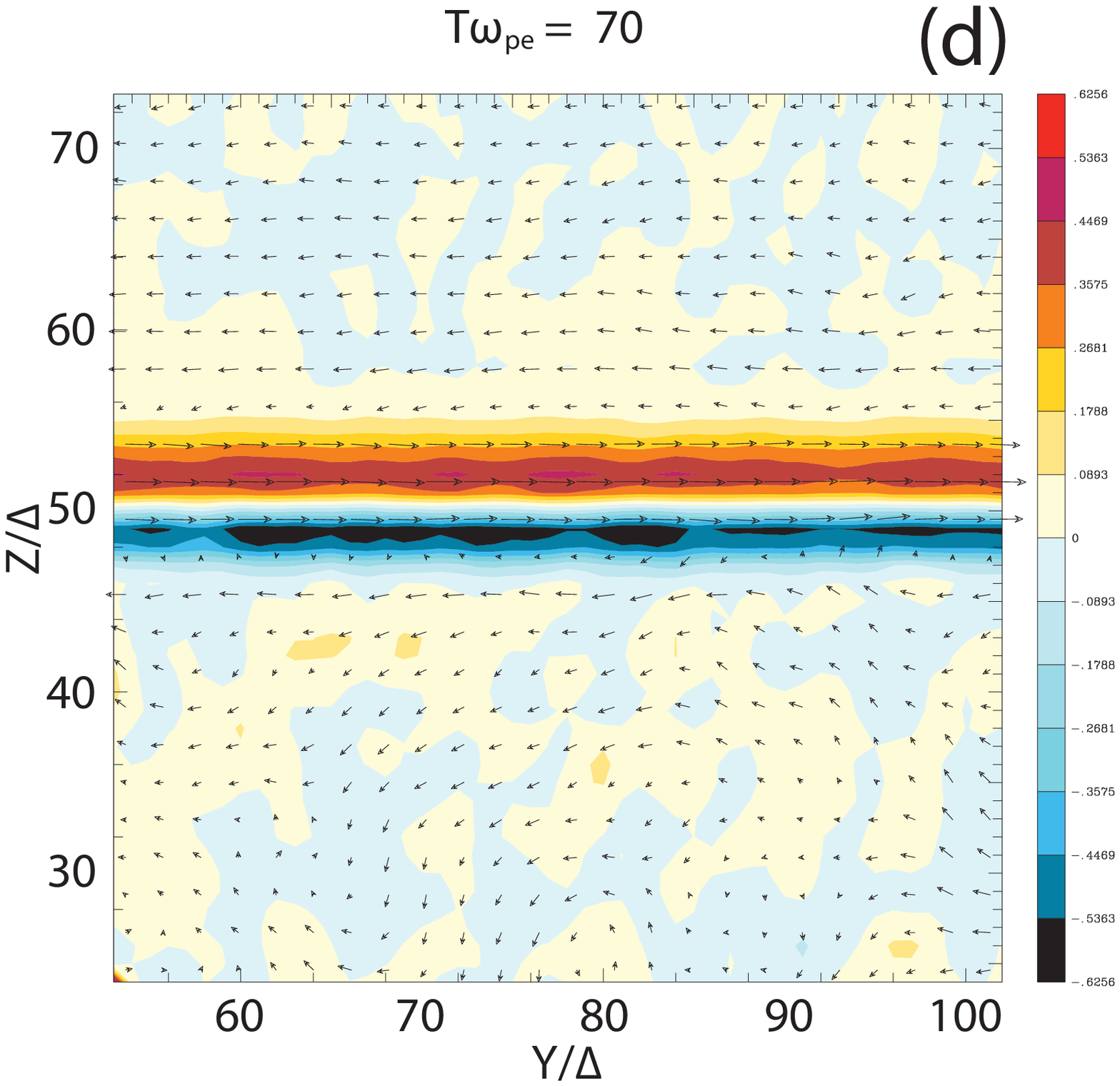}} 
\vspace*{-0.3cm}
\caption{\baselineskip 12.0pt Figure \ref{jxd} shows the time evolution of current filaments ($J_{\rm x}$) in the area 
denoted by small box in Fig \ref{kkhi}c.   KKHI starts to grow at $t = 30 \omega_{\rm pe}^{-1}$ (a) and current filaments have
merged by  $t = 70 \omega_{\rm pe}^{-1}$ (d). The maximum current density (simulation units) is (a) $\pm0.105$ at $t = 30 \omega_{\rm pe}^{-1}$, (b) $\pm 0.167$ at $t = 40 \omega_{\rm pe}^{-1}$, (c) $\pm 0.296$ at $t = 50 \omega_{\rm pe}^{-1}$, and (d) $\pm 0.625$
at $t = 70 \omega_{\rm pe}^{-1}$. The arrows show the magnetic field ($B_{\rm y}, B_{\rm z}$) (the length of the arrows
are not scaled to the strength of the magnetic fields). 
\label{jxd}}
\end{figure*}

\vspace{-0.cm}
\subsection{Our New Core-Sheath Jet KKHI Results}
\vspace{-0.cm}

The simulation setup for our study of velocity shears (not  counter-streaming shear flows as used by Alves et al. [2012] and Liang et al. [2012]) is shown in Fig. \ref{khiI}c.
In our simulation a relativistic jet  plasma is surrounded by a sheath plasma.  This setup is similar to the setup of our
RMHD simulations \citep{mizuno07}. In our initial simulation the jet core has $v_{\rm core} = 0.9978c \,(\gamma=15)$ pointing in the positive $x$ direction in the middle of the simulation box as in Alves et al. 2012. Unlike Alves et al.\ the upper and lower quarter of the simulation box contain a stationary, $v_{\rm sheath}= 0$, sheath plasma. Our setup allows for motion of the sheath plasma in the positive $x$ direction. 

Overall, this structure is similar in spirit, although not in scale, to that proposed for active
galactic nuclei (AGN) relativistic jet cores surrounded by a slower moving sheath,
and is also relevant to gamma-ray burst (GRB) jets. In particular, we note that
this structure is also relevant to the â jet-in-a-jetâ or needles in a jet scenarios
\citep[and papers therein]{giannios09}, which have been invoked to provide smaller
scale high speed structures within a much larger more slowly moving AGN jet.
Similar smaller scale structures within GRB jets are also conceivable.

This more realistic setup is different from the initial conditions used by the
previous simulations with counter-steaming flows \cite{alves12}, and hence allows
us to compute synthetic spectra in the observer frame.
As mentioned by Alves et al. [2012], in a non-counterstreaming or unequal density counterstreaming setup the growing kinetic KHI will propagate
with the flow. For GRB jets, the relativistic jet core will have much higher density
relative to the external medium. On the other hand, for an AGN jet the relativistic
core is less dense than the surrounding sheath.

We have performed a simulation using a system with $(L_x , L_y , L_z ) = (1005\Delta, 205\Delta, 205\Delta)$ and with an ion to electron mass ratio of  $m_{\rm i}/m_{\rm e} = 20$. Figure \ref{kkhi} shows the magnetic field structures generated by the relativistic electron-ion core with $\gamma = 15$ and with a stationary sheath plasma at time $t = 70 \, \omega_{\rm pe}^{-1}$.  Figure \ref{kkhi}a shows the magnitude of $B_{\rm y}$  plotted in the $y - z$ plane (jet flow is out of the page) at the center of the simulation box, $x = 500\Delta$. Figure 3d shows the magntude of $B_{\rm y}$ in the $x - z$ plane (jet flow in the $+x$-direction indicated by the arrow) at the center of the simulation box, $y = 100\Delta$. Figure 3b  shows $B_{\rm y}$ (red), $B_{\rm x}$ (black), and $B_{\rm z}$ (blue) magnetic field components at $x = 500\Delta$ and $y = 100\Delta$. Figure \ref{kkhi}d shows the $x$ component of the current. Relativistic jet flow is out of the page and positive (red) current flows along the jet side, whereas negative (blue) current flows along the sheath side. Positive currents are stronger than the negative currents, leading to $B_{\rm y}$ as shown in Figs. \ref{kkhi}a and \ref{kkhi}d. In forthcoming work we will obtain synthetic spectra from particles accelerated by KKHI as we have done for shock simulations \citep{nishikawa11,nishikawa12}.

\section{Summary and Discussion}  

Fig. \ref{jxd} shows KKHI growth, current filament merger and strong magnetic field generation in our simulation.
The structure of  KKHI seems similar to that found in counter-streaming simulations.
In order to examine growth rates and wave propagation, we have rederived the linear theory to describe our core-sheath jets.
 
We have extended the analysis presented in Gruzinov (2008) to core-sheath electron-proton plasma flows allowing for different core and sheath densities $n_{\rm jt}$ and $n_{\rm am}$, respectively, and core and sheath velocities $v_{\rm jt}$ and $v_{\rm am}$, respectively. In this analysis the protons are considered to be free-streaming whereas the electron fluid quantities and fields are linearly perturbed.
The dispersion relation becomes:

\begin{eqnarray}
& &(k^{2}c^{2} + \gamma^{2}_{\rm am}\omega^{2}_{\rm p,am} - \omega^{2})^{1/2} (\omega - kV_{\rm am})^{2} \nonumber \\
& & \times [(\omega - kV_{\rm jt})^{2} - \omega^{2}_{\rm p,jt}] \nonumber \\
&  +& (k^{2}c^{2} + \gamma^{2}_{\rm jt}\omega^{2}_{\rm p,jt} - \omega^{2})^{1/2} (\omega - kV_{\rm jt})^{2} \nonumber \\
 & & \times [(\omega - kV_{\rm am})^{2} - \omega^{2}_{\rm p,am}]  = 0 ,
\end{eqnarray}
where $\omega_{\rm p,jt}$ and  $\omega_{\rm p,am}$ are the plasma frequencies ($\omega^{2}_{\rm p} \equiv 4\pi n e^{2}/\gamma^{3}m$) of jet and ambient electrons, respectively, $k$
is the wave number parallel to the jet flow, and $\gamma_{\rm jt}$ and  $\gamma_{\rm am}$  are Lorentz factors of jet and ambient electrons, respectively.

Analytic solutions are not available except in the low ($\omega \ll \omega_{\rm p}$ and $kc \ll \omega_{\rm p}$) and high 
($\omega \gg \omega_{\rm p}$ and $kc \gg \omega_{\rm p}$)  frequency and wavenumber limits.  Equation (2) provides an analytic solution to the dispersion relation in the small wavenumber limit.  Here the real part gives the phase velocity and the imaginary part gives the temporal growth
rate and directly shows the dependence of the growth rate on the velocity difference across the
shear surface.

\begin{eqnarray}
\omega & \sim & \frac{(\gamma_{\rm jt}\omega_{\rm p,am}/\omega_{\rm p,jt})}{
(1 + \gamma_{\rm jt}\omega_{\rm p,am}/\omega_{\rm p,jt})}
kV_{\rm jt}\nonumber \\
& \pm & i \frac{(\gamma_{\rm jt}\omega_{\rm p,am}/\omega_{\rm p,jt})^{1/2}}
{(1 + \gamma_{\rm jt}\omega_{\rm p,am}/\omega_{\rm p,jt})}
k(V_{\rm jt} - V_{\rm am}).
\end{eqnarray}
Equation 2 shows that the wave speed increases and the temporal growth rate decreases as
$\gamma_{\rm jt}\omega_{\rm p,am}/\omega_{\rm p,jt} = \gamma^{5/2}
_{\rm jt} (n_{\rm am}/n_{\rm jt})^{1/2}$ increases. 
Numerical solution of the dispersion relation will be used for comparison with growth rates and wave propagation speeds observed in future simulations.

\begin{acknowledgments}
This work is supported by NSF AST-0908010, and AST-0908040, NASA-NNG05GK73G,
NNX07AJ88G, NNX08AG83G, NNX08 AL39G,  NNX09AD16G, and NNX12AH06G. JN is supported by NCN through grant DEC-2011/01/B/ST9/03183.  Simulations were performed at the Columbia and Pleiades facilities at the
NASA Advanced Supercomputing (NAS) and Kraken and Nautilus at The National Institute for Computational Sciences (NICS) which is supported by the NSF. This research was started during the program ``Chirps, Mergers and Explosions:  The Final Moments of Coalescing Compact Binaries'' at the Kavli Institute for Theoretical Physics which is supported by  the National Science Foundation under Grant No. PHY05-51164.
\end{acknowledgments}

\end{document}